\documentclass[letterpaper, final, 10 pt, conference]{ieeeconf}  

\makeatletter
\let\NAT@parse\undefined
\makeatother
\usepackage{cite}
\usepackage{float}
\usepackage{subfig}
\usepackage{amsfonts}

\IEEEoverridecommandlockouts                             

\usepackage{graphicx} 
\usepackage{epsfig} 
\usepackage{mathptmx} 
\usepackage{times} 
\usepackage{amsmath} 
\usepackage{amssymb}  
\usepackage{siunitx}
\usepackage{caption}
\usepackage{epstopdf}
\title{\LARGE \bf
No More Differentiator in PID:\\ Development of Nonlinear Lead for Precision Mechatronics
}

\author{Arun Palanikumar, Niranjan Saikumar, S. Hassan HosseinNia  
\thanks{The authors are with Department of Precision
	and Microsystems Engineering, Delft University of Technology, Delft, The
	Netherlands}
\thanks{{\tt\small A.Palanikumar@student.tudelft.nl,}}
\thanks{{\tt\small N.Saikumar@tudelft.nl,}}
\thanks{{\tt\small S.H.HosseinNiaKani@tudelft.nl}}%
}

\begin{document}
	
\maketitle

\begin{abstract}

Industrial PID consists of three elements: Lag (integrator), Lead (Differentiator) and Low Pass Filters (LPF). PID being a linear control method is inherently bounded by the waterbed effect due to which there exists a trade-off between precision \& tracking, provided by Lag and LPF on one side and stability \& robustness, provided by Lead on the other side. Nonlinear reset strategies  applied in Lag and LPF elements have been very effective in reducing this trade-off. However, there is lack of study in developing a reset Lead element.  In this paper, we develop a novel lead element which provides higher precision and stability compared to the linear lead filter and can be used as a replacement for the same. The concept is presented and validated on a Lorentz-actuated nanometer precision stage. Improvements in precision, tracking and bandwidth are shown through two separate designs. Performance is validated in both time and frequency domain to ensure that phase margin achieved on the practical setup matches design theories.

\end{abstract}

\section{INTRODUCTION}
\label{sec:intro}
The high tech industry is at the forefront of pushing the limitations and barriers in motion control technology. Demands for achieving higher precision and speeds are ever increasing. The wafer scanner industry, which is involved in manufacturing integrated circuits, is a prime example where (sub) nanometer precision positioning is required, while at the same time meeting challenging throughput demands. For accurate and fast servo-positioning of mechanical actuators in real life engineering systems, high quality motion control is required.

PID and related linear controllers (includes use of other linear filters) have been the standard for industrial motion control for many years owing to various factors such as its wide applicability and ease of implementation. PID does not require a precise model or thorough knowledge of the system and can instead be tuned using standard guidelines.  With multiple advancements in feedforward control techniques, it has become possible to achieve high bandwidths and precision using PID. However, when extreme demands have to be met, the fundamental limitations of linear control, to which PID is not exempt, become more evident. Such inherent limitations cause different performance criteria to be conflicting in nature, making it impossible to improve one criterion without negatively influencing another. This phenomenon is known in linear control theory as waterbed effect. For a more detailed explanation of the water bed effect, \cite{RCS} and \cite{waterbed} can be referred.
 
Requirements for precision motion control have been discussed in detail in \cite{munnig_shmidt} and \cite{precision}. Loop shaping, one of the popular methods for designing PID controllers in the high tech industry, is a technique in which the controller is designed in such a way that the frequency response of the open loop transfer function has the desired shape in gain and phase. For good tracking, high gain at frequencies upto the bandwidth is required. In PID, the integrator plays the role of increasing gain at low frequencies. Simultaneously, gain at high frequencies needs to be low to effectively attenuate noise in the system and hence provide good precision. This is commonly achieved through the use of low pass filters. In the frequency region around the bandwidth, phase behavior determines the stability and robustness of the system 	. Phase addition in PID is achieved by the tamed differentiator or lead component. The constraint in the relation between gain and phase behavior in linear systems is explained by Bode's gain-phase relationship \cite{lim1}. Hence attempts to change gain behaviour to improve precision and/or tracking negatively affects phase and hence stability and vice-versa. Also attempts to improve bandwidth results in decrease in phase margin and deterioration in precision performance. These limitations are fundamental to linear controllers including advanced optimal linear control and can only be overcome with the use of nonlinear control.

Reset control is one such promising technique which can be used for this purpose. Reset control was first introduced by J.C. Clegg in 1958, in the form of an integrator whose state is reset whenever the input error signal crosses zero \cite{clegg}. Approximation of the behavior of Clegg integrator (CI) through describing function analysis shows that the CI has a phase lag of about $\SI{-38}{\degree}$, which is $\SI{52}{\degree}$ higher than that of a linear integrator, for similar gain behavior. Since then, a lot of work has been done on reset control theory and the benefits of using reset controllers in various practical applications have been investigated. For example, reset control has been used to achieve improved performance in hard-disk-drive systems,\cite{hdd},\cite{hdd2} and mechatronic systems \cite{hassan1},\cite{hassan2}. In \cite{pzt}, it is shown that reset can be used to reduce overshoot and improve settling time in PZT positioning stages. Despite the promise shown, it is observed that much of the research in reset control was focused on the integrator. Some efforts have been made in investigating other reset elements as well. For example, in \cite{gfrore}, a generalized fractional order reset element is introduced. In \cite{lag}, a reset lead-lag filter has been used to improve mid-frequency disturbance rejection. And in \cite{sore_stage}, a second order reset element has been introduced to be used in stage control design. However apart from reset integrator, the use of reset elements within the framework of PID for improved performance has not been well investigated.

The motivation for this research arises from the possibility of reducing the severity of the waterbed effect limitation in PID using reset control. In this paper, reset has been implemented in the taming pole of the lead component in PID, and a Reset PID controller has been developed using common loop shaping techniques. The main contribution of this work is the concept of reset PID to overcome fundamental limitations of linear control along with two cases of reset PID controller design with first design used to improve tracking and precision for same bandwidth and second design for increase in bandwidth and tracking for same precision.

In section \ref{sec:preliminaries}, necessary background information about reset control is provided. The concept of reset PID is provided in section \ref{sec:resetpid}. The experimental setup used for validation of the developed controllers is described in section \ref{sec:setup}. In section \ref{sec:results}, experimental results from implementation of the two controllers are given, and compared with the performance of linear PID controller, followed by conclusions.
 
\section{PRELIMINARIES}
  \label{sec:preliminaries}

\subsection{Definition of Reset Control}
\begin{figure}
	\centering
	\includegraphics[width=1\linewidth]{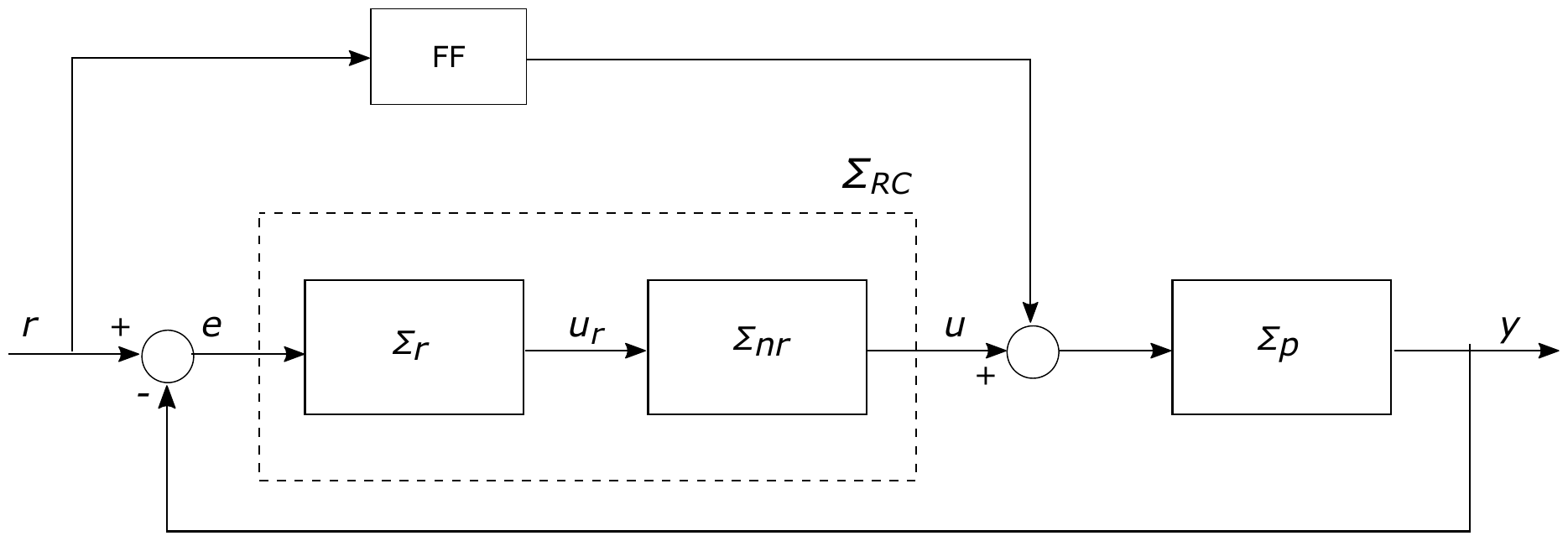}
	\caption{Block diagram of feedback loop with a reset controller $\Sigma_{RC}$, and plant $\Sigma_{p}$. Feedforward term FF is calculated as given in \cite{profile4}.}
	\label{fig:block_reset}
\end{figure}

A general reset controller can be defined as follows, using the notations used in \cite{RCS}: 
\begin{equation}\label{eq:rcs}
\Sigma_{RC}:=\begin{cases}
\dot{x}_r(t)=A_Rx_r(t)+B_Re(t)&\text{if } e(t)\neq0,\\
x_r(t^+)=A_\rho x_r(t)&\text{if } e(t)=0\\
u(t)=C_Rx_r(t)+D_Re(t)& \\
\end{cases}
\end{equation}
where matrices $A_R, B_R, C_R, D_R$ are the base linear state-space matrices of the control system. $A_\rho$ is the reset matrix, which determines the after reset values of the states. $u(t)$ is the 
control input and the error signal $e(t)$ is the difference between reference $r(t)$ and output $y(t)$. $x_r(t)$ denotes the controller states. A typical control structure using reset control is shown in Fig. \ref{fig:block_reset}. The reset controller $\Sigma_{RC}$ is separated into two components : a nonlinear part $\Sigma_{r}$ whose states are reset, and a linear part $\Sigma_{nr}$ whose states are not reset. The two components are connected in series, with the output $u_r(t)$ from reset part given as input to the non reset part.

\subsection{Describing Function Analysis}

Frequency response of reset elements can be approximated using describing function analysis. This method makes the assumption that non linear system behavior has a quasi-linear amplitude dependent relation between sinusoidal excitation inputs and the steady state response at the fundamental excitation frequency. The higher harmonics are neglected under the assumption that they are sufficiently filtered out by the low pass characteristics of the system. Describing functions thus help extend the use of loop shaping techniques to reset control. For a more detailed explanation of describing functions, \cite{DF} can be referred.

The general describing function of a reset system as defined in \cite{hdd} is given by:
\begin{equation}\label{eq:DF}
G_\mathrm{DF}(j\omega)=C_R(j\omega I-A_R)^{-1}B_R(I+j\Theta_D(\omega))+D_R
\end{equation}
with
\begin{equation}\label{eq:thetad}
\Theta_D(\omega)=-\frac{2\omega^2}{\pi}\Delta(\omega)[\Gamma_D(\omega)-\Lambda^{-1}(\omega)]
\end{equation}
where the following equations have been used:
\begin{eqnarray*}
	\begin{cases}
		\Lambda(\omega)=\omega^2I+A^2_R\\
		\Delta(\omega)=I+e^{\frac{\pi}{\omega}A_R}\\
		\Delta_D(\omega)=I+A_\rho e^{\frac{\pi}{\omega}A_R}
	\end{cases}
\end{eqnarray*}

\subsection{Controlling nonlinearity with $\gamma$}

The reset matrix $A_\rho$ in \eqref{eq:rcs} provides a degree of freedom in tuning the system and is defined as follows: 
\begin{equation}
 	A_\rho = \left[ \begin{array}{cc}
		 \gamma I_{n_r} & 0 \\0 & I_{n_{nr}}
 	\end{array}  \right]
\end{equation} 
where $n_r $ and $n_{nr}$ are the number of states in $ \Sigma_r $ and  $ \Sigma_{nr} $ respectively.  
 
\begin{figure}[H]
	\centering
	\includegraphics[width=\linewidth]{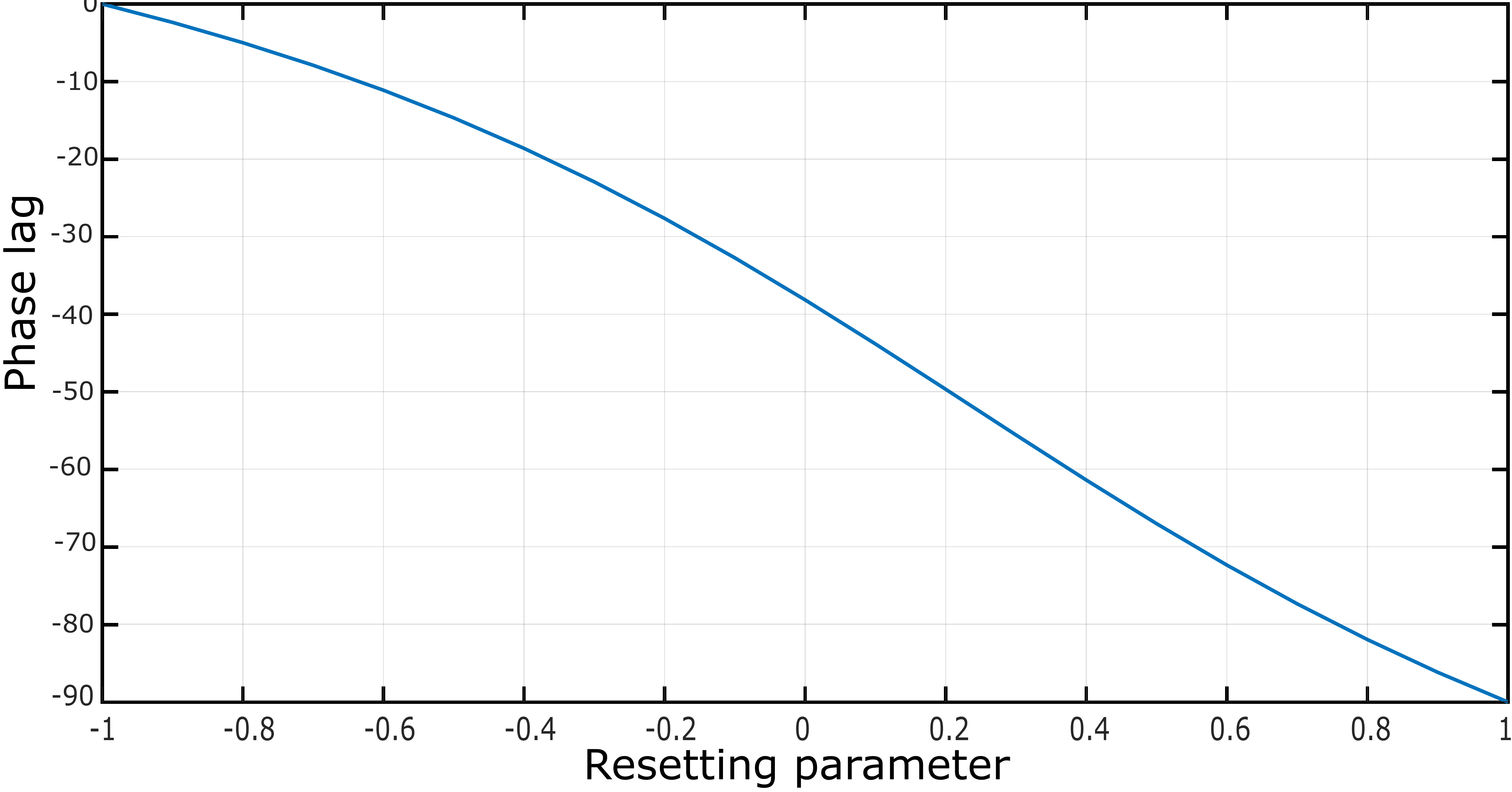}
	\caption{Variation of phase lag with change in $\gamma$.}
	\label{fig:gfore_phase}
\end{figure}

If the resetting parameter $\gamma = 1$, the reset element reduces to its base linear system as no reset occurs and $\gamma = -1$ denotes extreme reset. As $| \gamma - 1| $ increases, the nonlinearity in the system increases because the magnitude of jump in state value increases. Increase in nonlinearity is not desired due to the associated higher order dynamics. The variation of phase lag with change in $\gamma$ is shown in Fig. \ref{fig:gfore_phase}, and it can be seen that the more nonlinear the system is, greater is the phase benefit.


\begin{figure}[H]
 	\centering
 	\includegraphics[width=\linewidth]{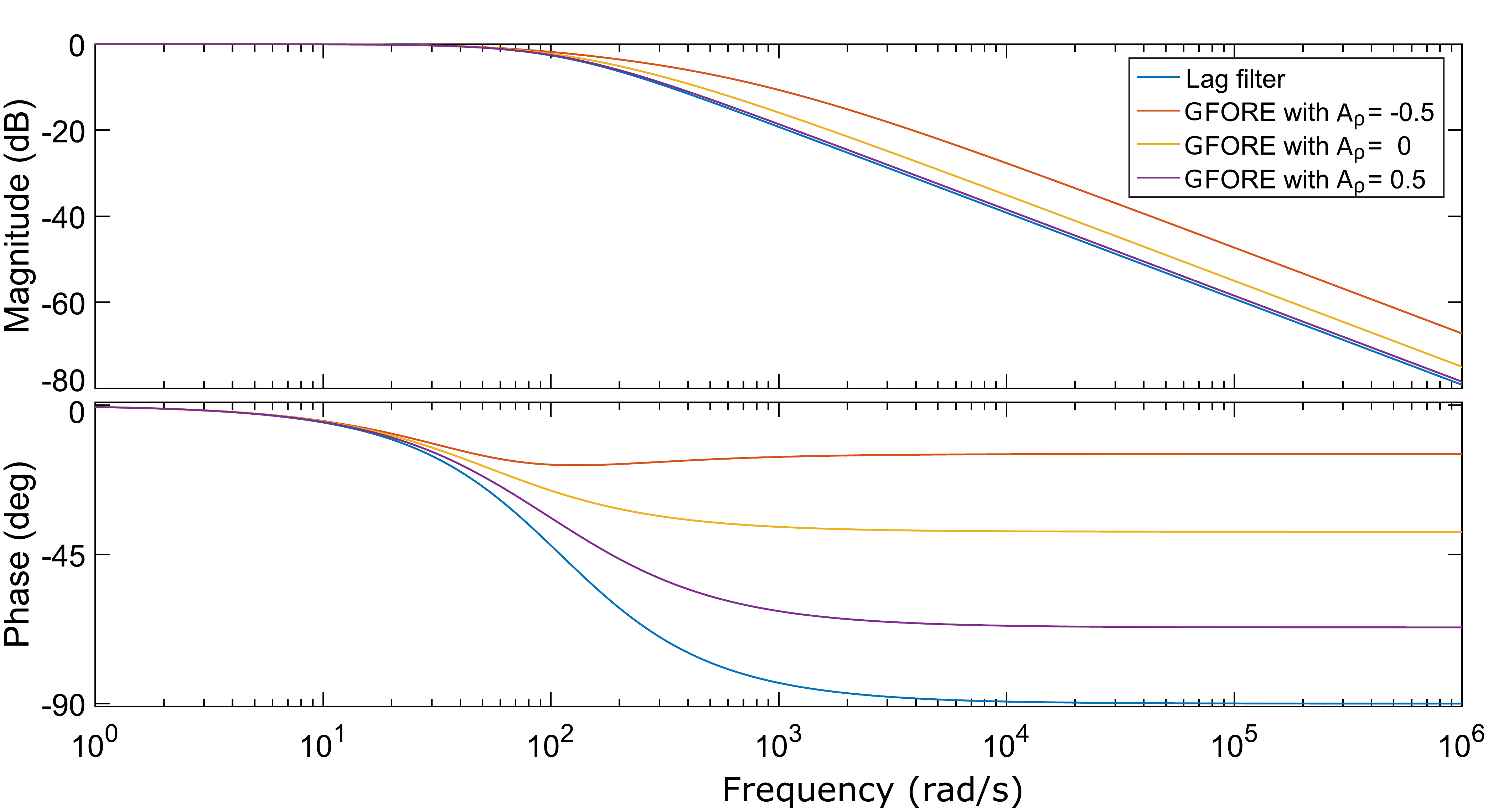}
 	\caption{Frequency response of FORE compared with that of a linear lag  filter.}
 	\label{fig:fore_vs_lpf}
\end{figure}

Describing function of the generalized first order reset element (GFORE) developed in \cite{hdd}, is compared with a linear lag filter in Fig. \ref{fig:fore_vs_lpf}. It can be seen that for the case of $\gamma = 0$, upto the corner frequency at $\SI{100}{\Hz}$, the behavior of GFORE is similar to that of the lag filter. At higher frequencies, the phase lag in GFORE is only $\SI{38}{\degree}$ (for $A_\rho = 0$), as compared to $\SI{90}{\degree}$ in lag filter, while gain behaviors are almost similar. Describing functions have also been plotted for $\gamma = -0.5$ and $\gamma = 0.5$ to depict the variation in frequency response with changing $\gamma$.
 
\begin{figure}[H]
	\centering
	\includegraphics[width=0.85\linewidth]{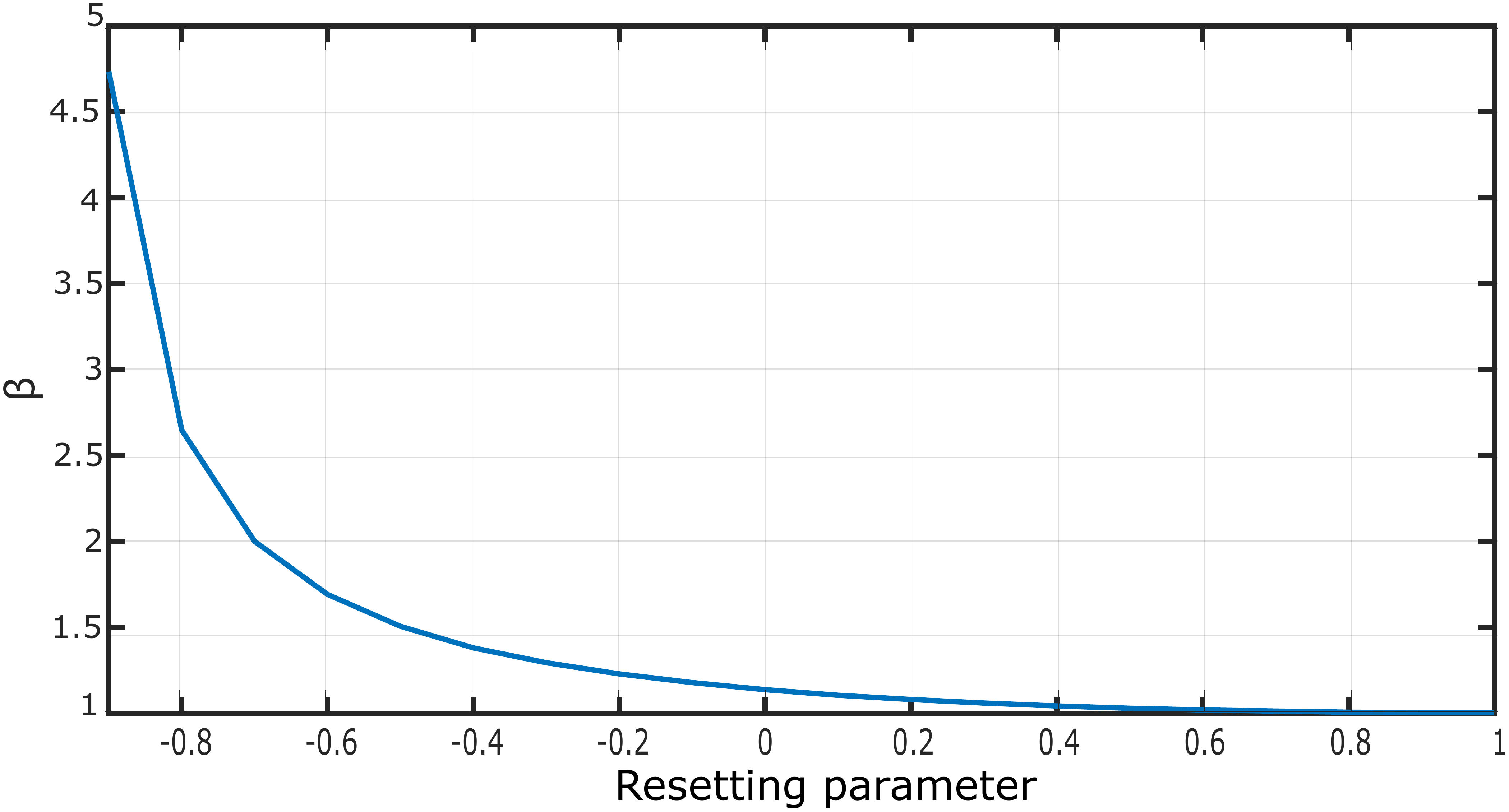}
	\caption{Variation of ratio $\beta$ with change in $\gamma$.}
	\label{fig:gfore_-3db}
\end{figure}

It can be seen that the cutoff frequency of GFORE also varies with change in $\gamma$ \cite{gfrore}. The ratio of cutoff frequency of GFORE to that of lag filter ($\beta$) is plotted in Fig. \ref{fig:gfore_-3db} for different values of  $\gamma$ . Choosing $\gamma$ values between -1 and 1 can help attain the desired trade-off between phase lag, cutoff frequency and non linearity.

\subsection{Stability Analysis}

The stability condition given in \cite{RCS} is used for
checking the stability of the system. The following condition was provided for ensuring asymptotic stability of a reset control system :

\textbf{Theorem}\cite{RCS}	{\it There exists a constant} $\beta\in \mathbb{R}^{n_{r}\times 1}$ {\it and} $P_{\rho}\in \mathbb{R}^{n_{r}\times n_{r}}, P_{\rho}>0$ {\it where} $n_{r}$ {\it is the number of reset states, such that the restricted Lyapunov equation} 
\begin{eqnarray}
P>0,\ A_{cl}^{T}P+PA_{cl}<0\\
B_{0}^{T}P=C_{0}
\end{eqnarray}  
{\it has a solution for} $P$, {\it where} $C_{0}$ {\it and} $B_{0}$ {\it are defined by}
 

\begin{align}
C_0=\left[\begin{array}{ccc}
\beta C_{nrp} & 0_{n_r \times n_{nr}} & P_\rho
\end{array}\right] , & &  B_0=\left[\begin{array}{c}
0_{n_{nrp} \times n_{r}}\\
0_{n_r \times n_{r}}\\
I_{n_r}
\end{array}\right]
\end{align}

 $A_{cl}$ is the closed loop matrix A-matrix
\begin{equation}
C_0=\left[\begin{array}{cc}
A_r & B_r C_{nrp}\\
-B_{nrp}C_r & A_{nrp}
\end{array}\right] 
\end{equation}
in which $(A_r,B_r,C_r,D_r)$ are the state space matrices of $\Sigma_r$ and 
$(A_{nrp},B_{nrp},C_{nrp},D_{nrp})$ are the state space matrices of $\Sigma_{nr}$ and $\Sigma_p$ in series.\\

This condition has been used to test stability of all developed controllers.

\section{RESET PID}
 \label{sec:resetpid}
Transfer function of linear PID in series form is given by:
\begin{equation}
{G_{PID}} ={K_{p}} \underbrace{\left(1+\frac{{\omega_i}}{s} \right)}_{\text{Lag component}} \underbrace{  \frac{\left( \frac{s}{\omega_d} +1  \right)}{\left( \frac{s}{\omega_t}  +1 \right)}}_{\text{Lead component}} \underbrace{\left(\frac{{1}}{\frac{s}{\omega_l}+1}  \right)}_{\text{LPF}}
\label{eq:pid1}
\end{equation}
where $\omega_i$ is the frequency at which integral action is stopped, and $\omega_l$ is the cut off frequency of the LPF. $K_{p}$ is the proportional gain. Differentiating action is started at $\omega_d$ and terminated at $\omega_t$, and therefore $\omega_t > \omega_d$.

\begin{figure}[!h]
	\centering
	\subfloat[]{\label{fig:theoryA_diff}\includegraphics[width=0.23\textwidth]{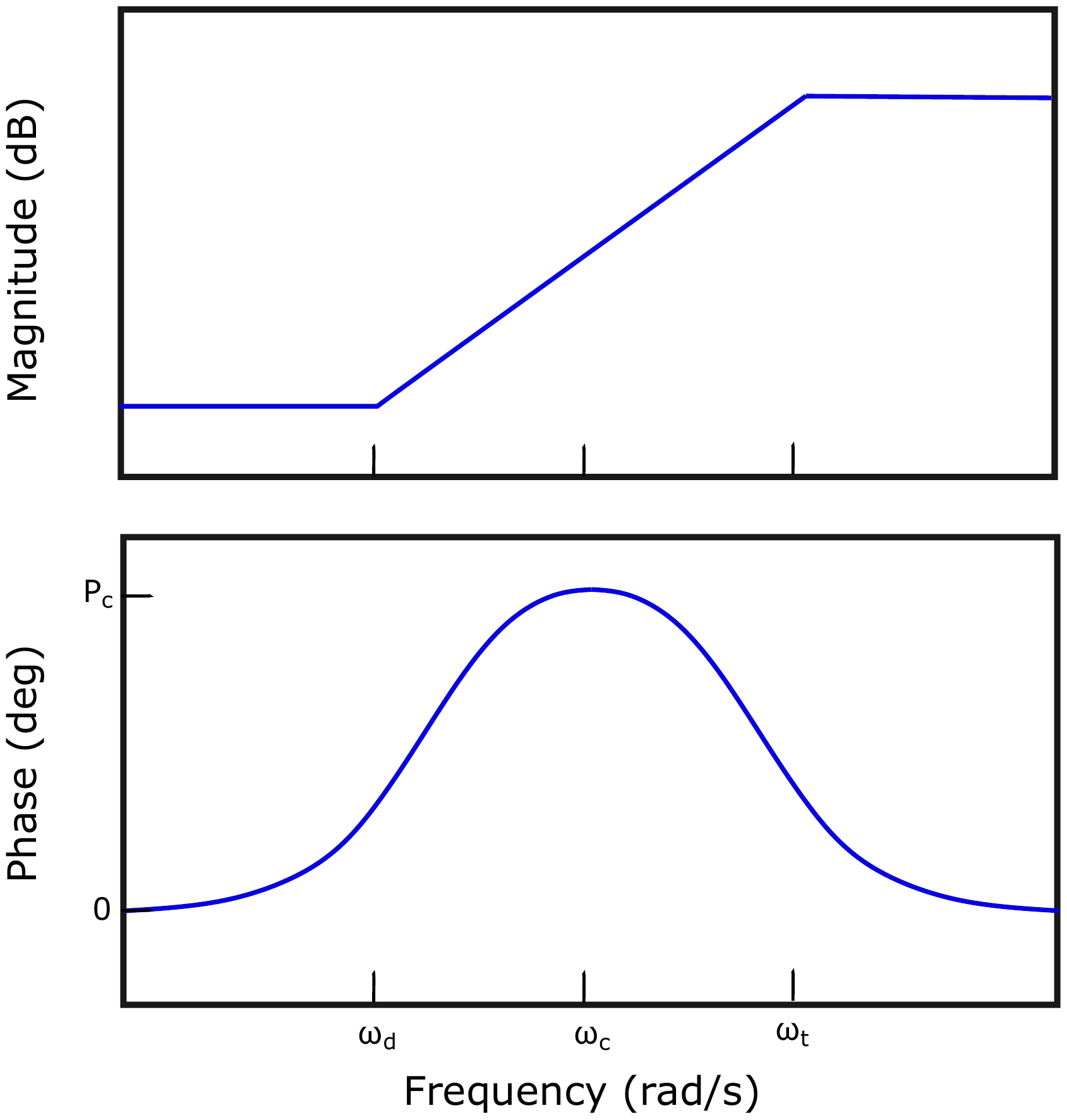}}
		\hspace{.1in}
	\subfloat[]{\label{fig:theoryA_diffcomp}\includegraphics[width=0.23\textwidth]{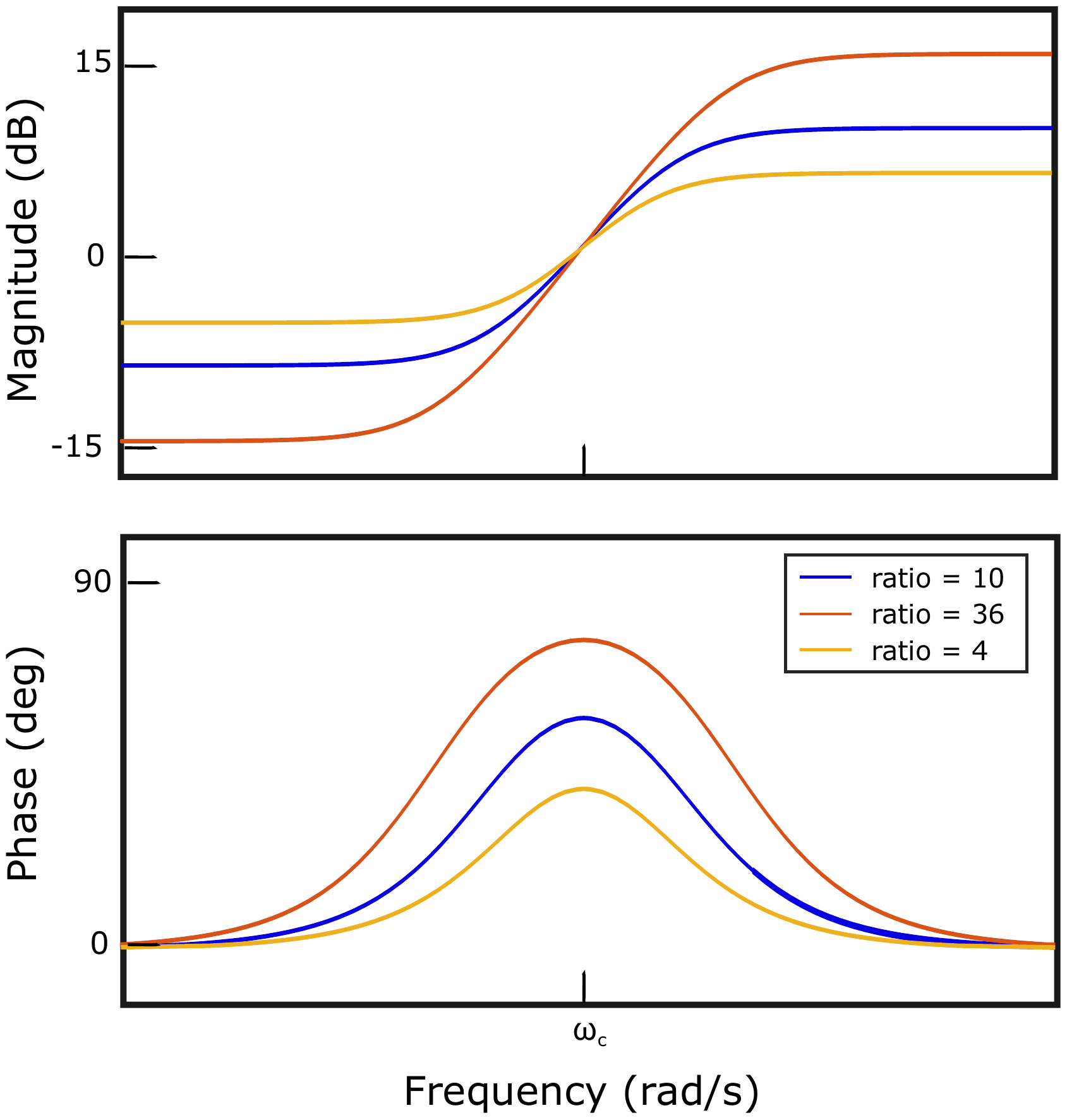}}
	\caption{\label{fig:theoryA_1} (a) Bode plot of a differentiator. (b) Bode plot showing increase in maximum phase and also high freqeuncy gain with increasing ratio of $\omega_t$ to $\omega_d$ }
\end{figure}

As discussed in section \ref{sec:intro}, differentiator or lead component of the PID adds essential phase to the system in bandwidth region to achieve stability and robustness. Phase at the crossover frequency also  affects the maximum overshoot of the system. Differentiator is tamed by placing a pole at $\omega_t$ because having large gains at higher frequencies is not desirable for noise attenuation and precision. The maximum phase lead that can be achieved from the differentiator in the absence of this taming pole would be $\SI{90}{\degree}$. However, the pole reduces the maximum phase lead to a value  which depends on frequency interval between the zero and the pole. By symmetrically placing the zero and pole around the crossover frequency $\omega_c$, it is ensured that the maximum phase from lead is achieved at bandwidth.  Fig. \ref{fig:theoryA_diff} shows the frequency response of a typical differentiator used in PID.

Higher and lower phase margins can be achieved by increasing or decreasing the frequency range between $\omega_d$ and $\omega_t$ respectively. The problem with this approach is that it results in a trade-off between tracking and precision performance on one side and stability and robustness on the other. This can be observed in Fig. \ref{fig:theoryA_diffcomp}. As the positive phase obtained increases with increase in the ratio of $\omega_t$ to $\omega_d$, the gain at higher frequencies is also increased, which affects the precision of the system. As a result, with the linear design, it is not possible to improve precision while having the same degree of robustness. Also, with increasing phase, there is a decrease in low frequency gain thereby affecting tracking behavior.

Reset can be introduced in this tamed differentiator to improve performance such that this fundamental limitation is overcome. Reset is introduced such that only the taming pole is reset resulting in the reset and non-reset parts given by:

\begin{equation}
\begin{array}{cc}
\Sigma_{nr} = \left( \frac{s}{\omega_d} +1  \right); & \Sigma_{r}=\frac{1}{\left( \frac{s}{\omega_t} +1  \right)  } 
\end{array}
\label{eq:tamedR}
\end{equation}

Although the non-resetting part as given in \eqref{eq:tamedR} is not a proper function, this in combination with integrator and LPFs will be proper. In the case that this is not true, an LPF can be added at a frequency where it does not affect phase at bandwidth, purely to make the function proper.

\begin{figure}[]
	\centering
	\subfloat[]{\label{fig:theoryA_tamed}\includegraphics[width=0.23\textwidth]{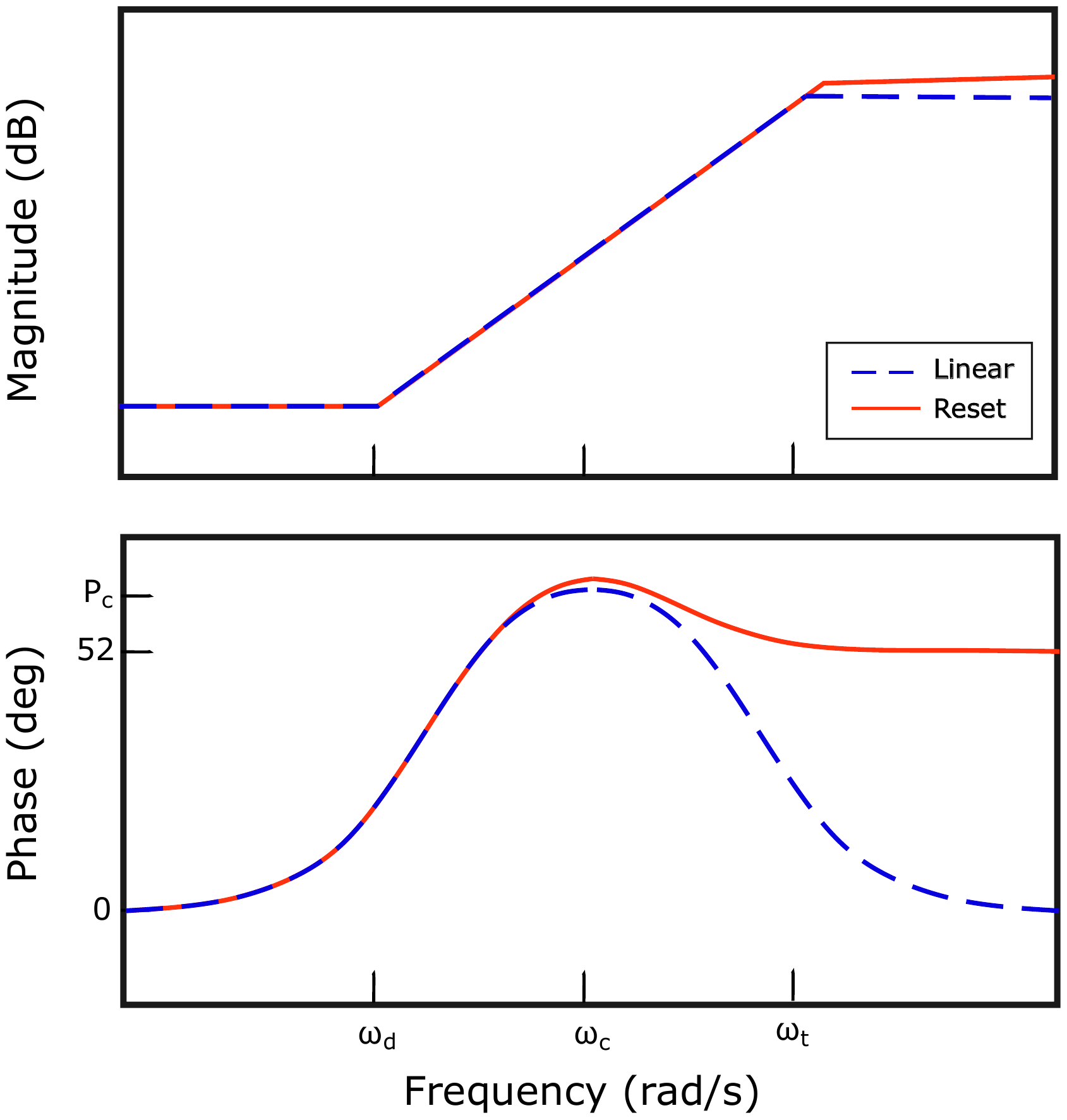}}
		\hspace{.1in}
	\subfloat[]{\label{fig:theoryA_diffvsll}\includegraphics[width=0.23\textwidth]{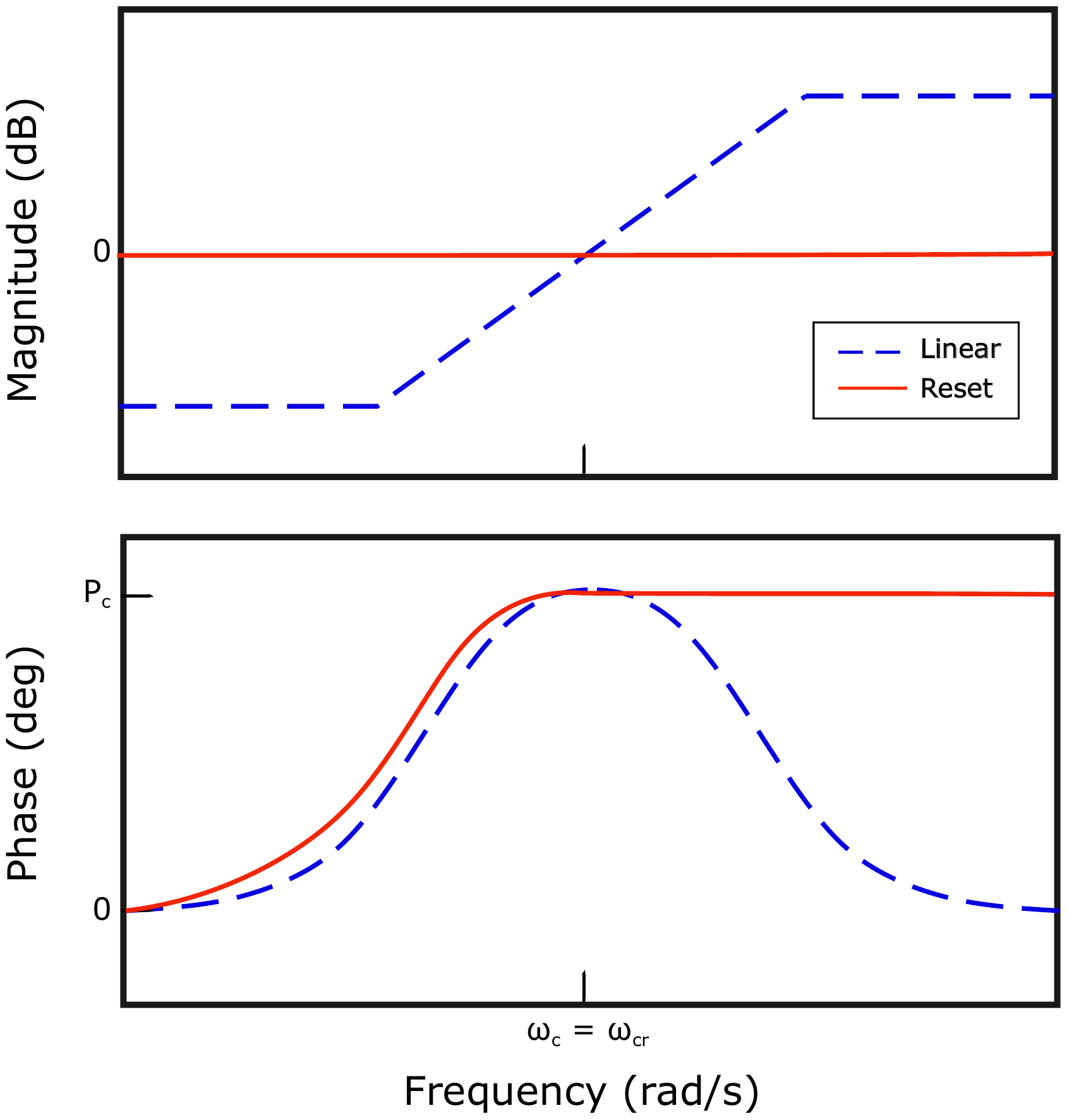}}
	\caption{\label{fig:theoryA_2}(a)Describing function analysis shows positive phase added by reset at frequencies above $\omega_t$ when compared to a linear lead component. (b) Reset can be used to achieve the same phase lead with favourable gain behavior.}
\end{figure}

The frequency response obtained through describing function analysis is compared with that of a linear tamed differentiator in Fig. \ref{fig:theoryA_tamed}. It can be seen that for a similar gain behavior, phase at frequencies higher than $\omega_t$ is not zero in the case of reset. If  $\gamma=0$ is used, the positive phase achieved would be 52 degrees (as shown in section \ref{sec:preliminaries}).
 
Although the resetting action adds very little phase around bandwidth compared to its linear counterpart, it creates phase lead at higher frequencies. This phase lead created can be moved towards the bandwidth region by shifting $\omega_t$ closer to $\omega_d$. If $\omega_t = \omega_d$, then complete phase lead generated will be through resetting action eliminating the linear lead filter and thus creating a $\SI{0}{\decibel}$ gain line with a positive phase, as shown in Fig. \ref{fig:theoryA_2}. To compensate for changes in cutoff frequency due to reset (as shown in Fig. \ref{fig:gfore_-3db}), a factor $\alpha$ is used. The reset element can be reformulated as given in \eqref{eq:ll}. $\gamma$ is chosen such that same phase margin is achieved. 

\begin{equation}
\label{eq:ll}
\Sigma_{r}=\frac{1}{\left( \frac{s}{\alpha \omega_d} +1  \right)  } 
\end{equation}

It can be observed that the gain at high frequency is lower than PID, while the low frequency gain is increased, without affecting the phase margin. It can be concluded that theoretically, better tracking and better precision has been achieved while maintaining the same robustness, which was impossible to do with linear control.

The reset element, when combined with an integrator and low pass filter, forms Reset PID controller A, given as

\begin{equation}
\Sigma_{RC} = \underbrace{\left( \frac{1}{\frac{s}{\alpha \omega_d}+1}   \right)}_{\Sigma_r} \underbrace{{K_{p}} \left(1+\frac{{\omega_I}}{s} \right) \left(1+\frac{s} {{\omega_d}} \right) \left(\frac{{1}}{\frac{s}{\omega_l}+1}  \right) }_{\Sigma_{nr}} 
\label{eq:rpid1}
\end{equation}

Although the resulting controller is termed `Reset PID', the traditional differentiator has been replaced by reset element which provides the required beneficial phase lead without the unnecessary gain increase. $K_{p}$ is chosen such that the open loop gain calculated from describing function analysis equals unity at $\omega_c$. 

\begin{figure}[!h]
	\centering
	\subfloat[ ]{\label{fig:theoryA_pidvsrpid}\includegraphics[width=0.23\textwidth]{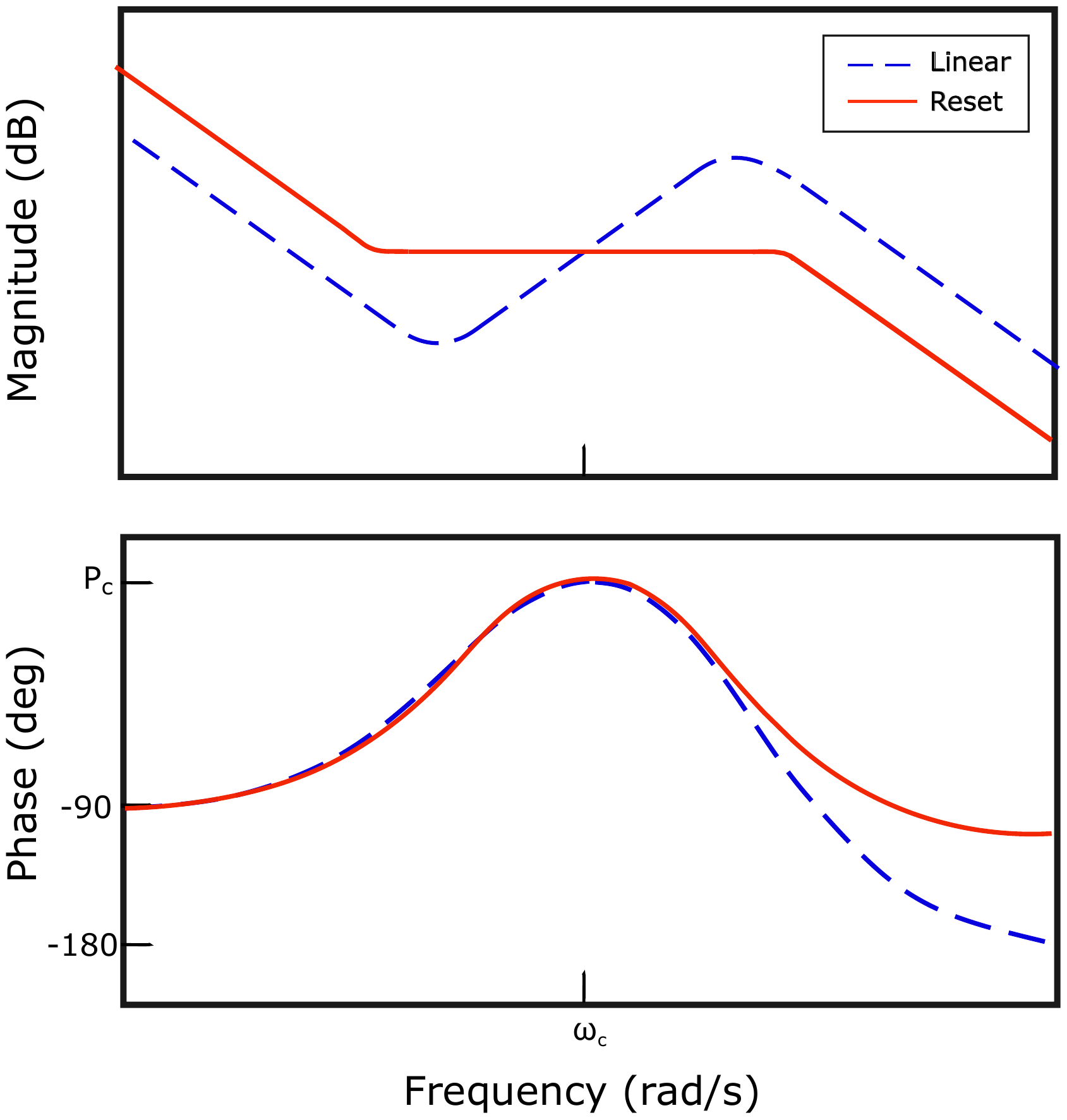}}
	\hspace{.1in}
	\subfloat[]{\label{fig:theoryB_pidvsrpid}\includegraphics[width=0.23\textwidth]{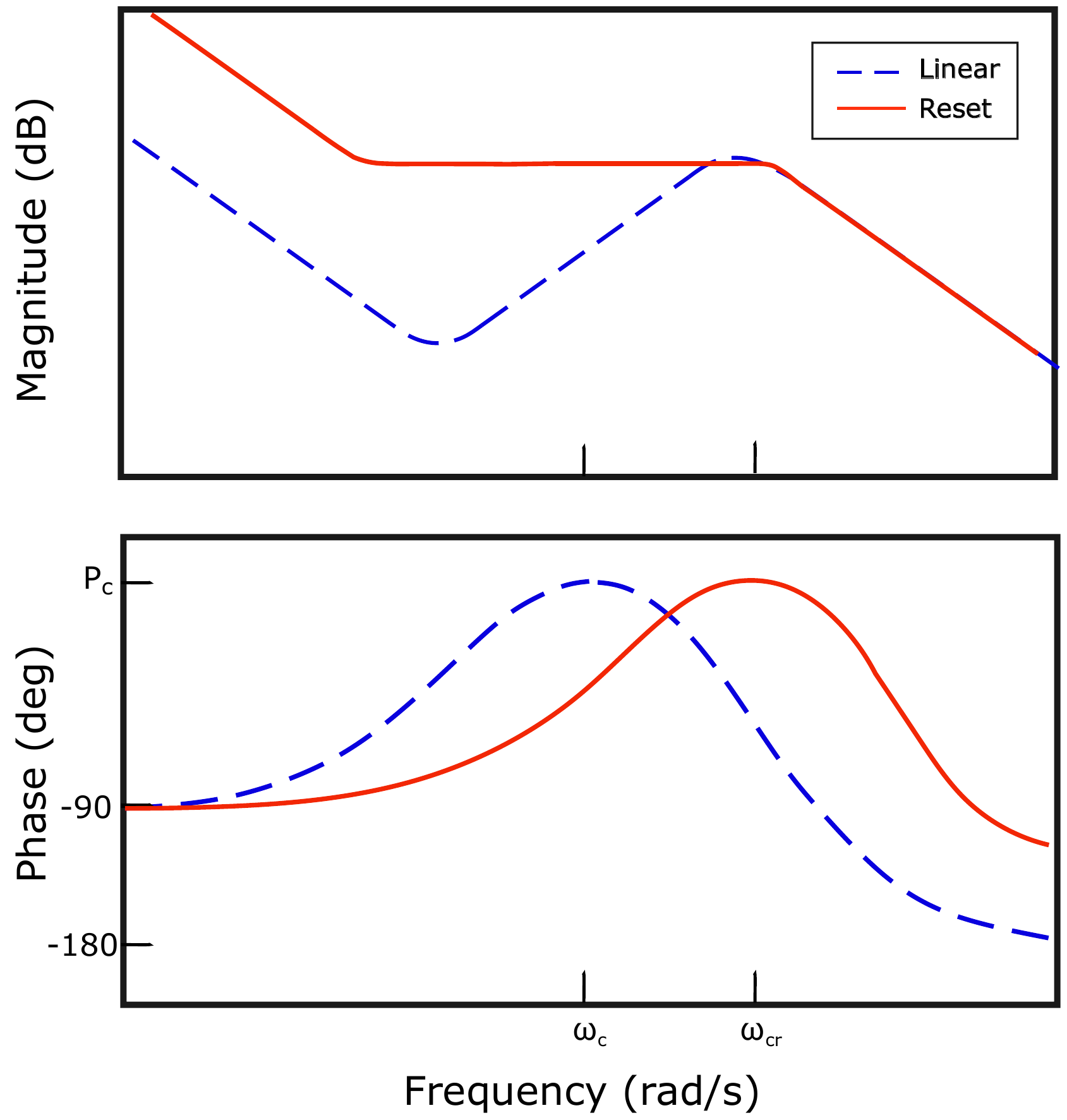}}
	\caption{\label{fig:theoryAB_3} (a) Comparison of describing function of Reset PID controller A with a PID controller having the same bandwidth and phase margin. (b) Comparison of describing function of Reset PID controller B with a PID controller having the same precision and phase margin. }
\end{figure}

By increasing the controller gain such that the gain at high frequencies equals that of PID, an increase in bandwidth can be achieved for the same phase margin and precision as shown in Fig. \ref{fig:theoryB_pidvsrpid}. It can be seen that bandwidth has increased from $\omega_c$ to $\omega_{cr}$. This controller, which provides improved tracking and bandwidth for the same phase margin and precision, will be referred to as Reset PID Controller B. Gain for Controller B at low frequencies is higher than that of Controller A, thereby indicating a further improvement in tracking performance. 

\section{EXPERIMENTAL SETUP}
 \label{sec:setup}
\begin{figure}[!h]
	\centering
	\subfloat[]{\label{fig:setup}\includegraphics[width=0.2\textwidth]{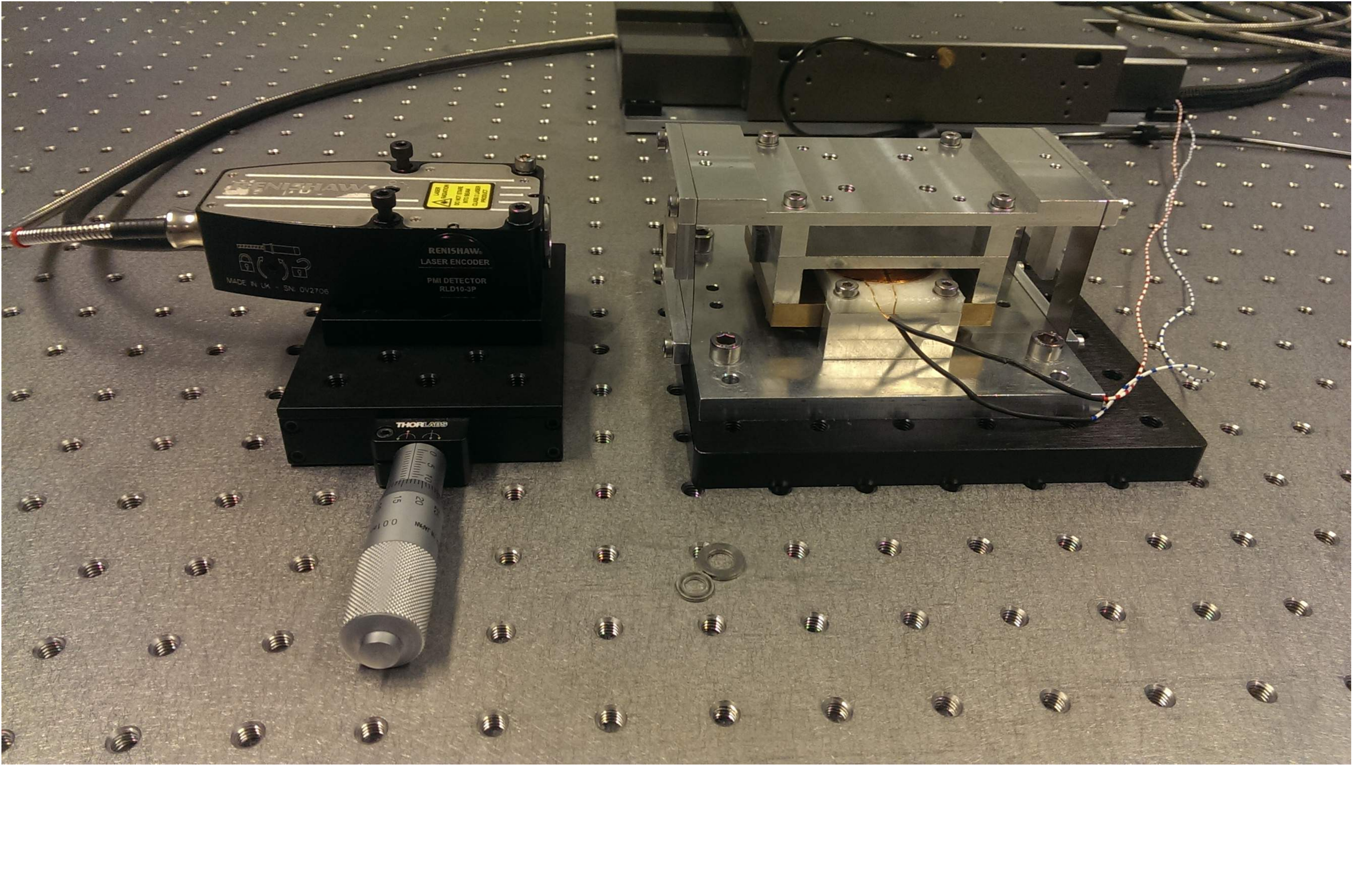}}
	\subfloat[]{\label{fig:bode_gfs}\includegraphics[width=0.28\textwidth]{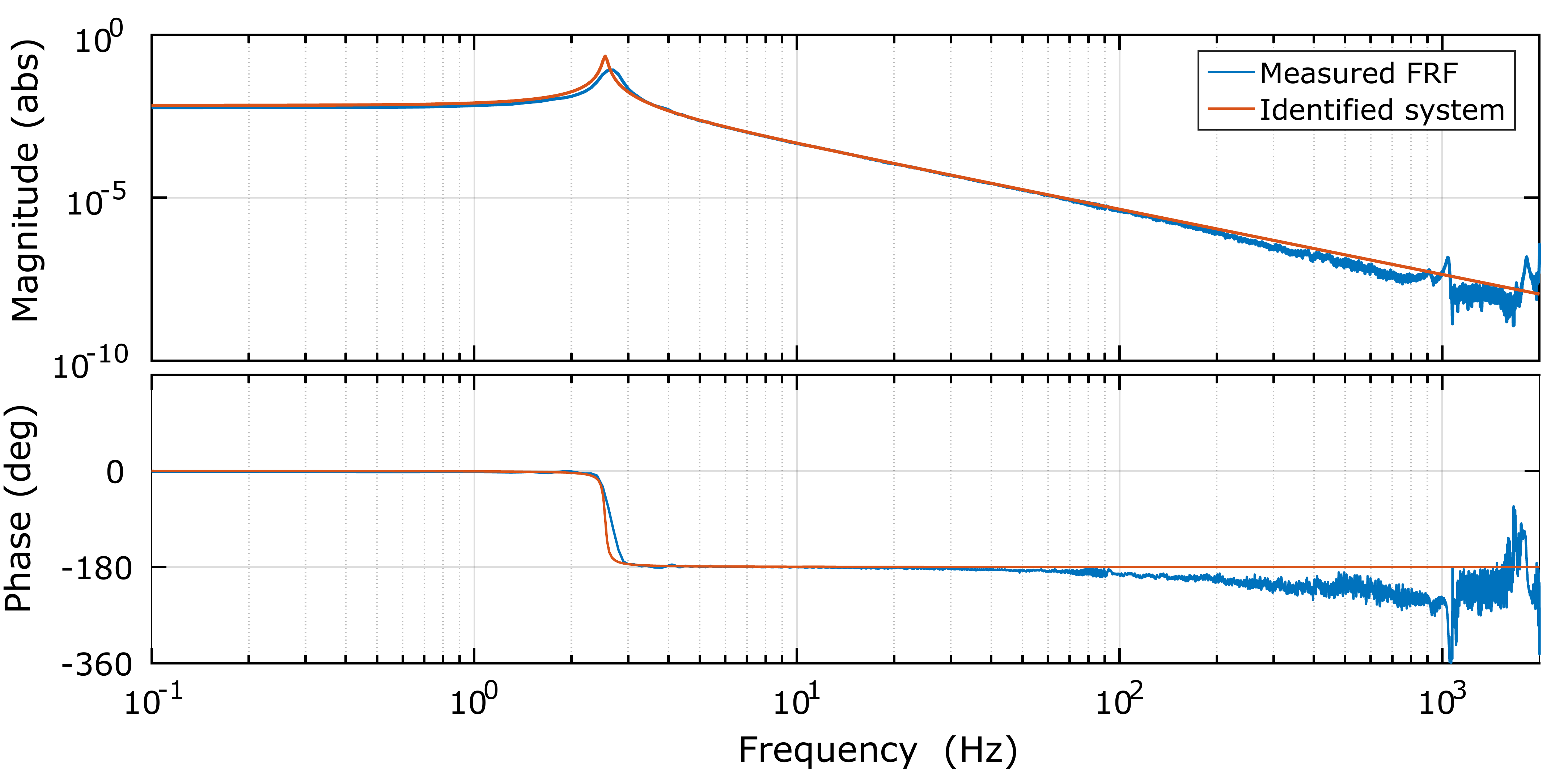}}
	\caption{(a) Picture of the Lorentz stage (right) with the laser encoder on the left.  (b) Frequency response of the system and
		the identified system model }
\end{figure}

The developed reset PID controller was implemented in the Lorrentz-actuated precision positioning stage shown in Fig. \ref{fig:setup}. The coil of the Lorentz actuator is attached to the stator and permanent magnets are mounted on the mover. Parallel leaf flexures are used for the linear guiding of the stage. The position of the stage is measured using a \textit{Renishaw RLE10} laser interferometer with a resolution of $\SI{10}{\nano\meter}$. Controllers are developed using MATLAB/Simulink environment and are run real time in a dSPACE DS1103 control system, which acts as the interface between the computers and the plant. A sampling rate of $\SI{20}{\kilo \Hz}$ is used for all controllers. 

The frequency response is shown in Fig. \ref{fig:bode_gfs} along with the identified system, which has the behavior of a second order mass-spring-damper system.

\section{EXPERIMENTAL VALIDATION}
 \label{sec:results}
 
\subsection{Controller Tuning}

This paper deals with the proposal and validation of reset PID and hence optimal tuning for performance is not presented. Instead controllers are heuristically designed to meet the required phase margin specifications. Controllers A and B are tuned for the precision stage, to have a bandwidth of $\SI{150}{\Hz}$ and $\SI{200}{\Hz}$ respectively, with a phase margin of $\SI{45}{\degree}$. $\omega_I$ is chosen to be $\omega_c / 10$, $\omega_d$ to be $\omega_c / 5$ and $\omega_l = \omega_c * 7$. $\alpha$ is taken as $0.7$. $K_p$ is $\num{9.24e5}$ for Controller A  and $\num{1.74e6}$ for Controller B.
 
\subsection{Time Domain Experiments}
\subsubsection{Reference tracking}
\begin{figure}[!h]
	\subfloat[]{\label{tr:a}\includegraphics[width=0.2395\textwidth]{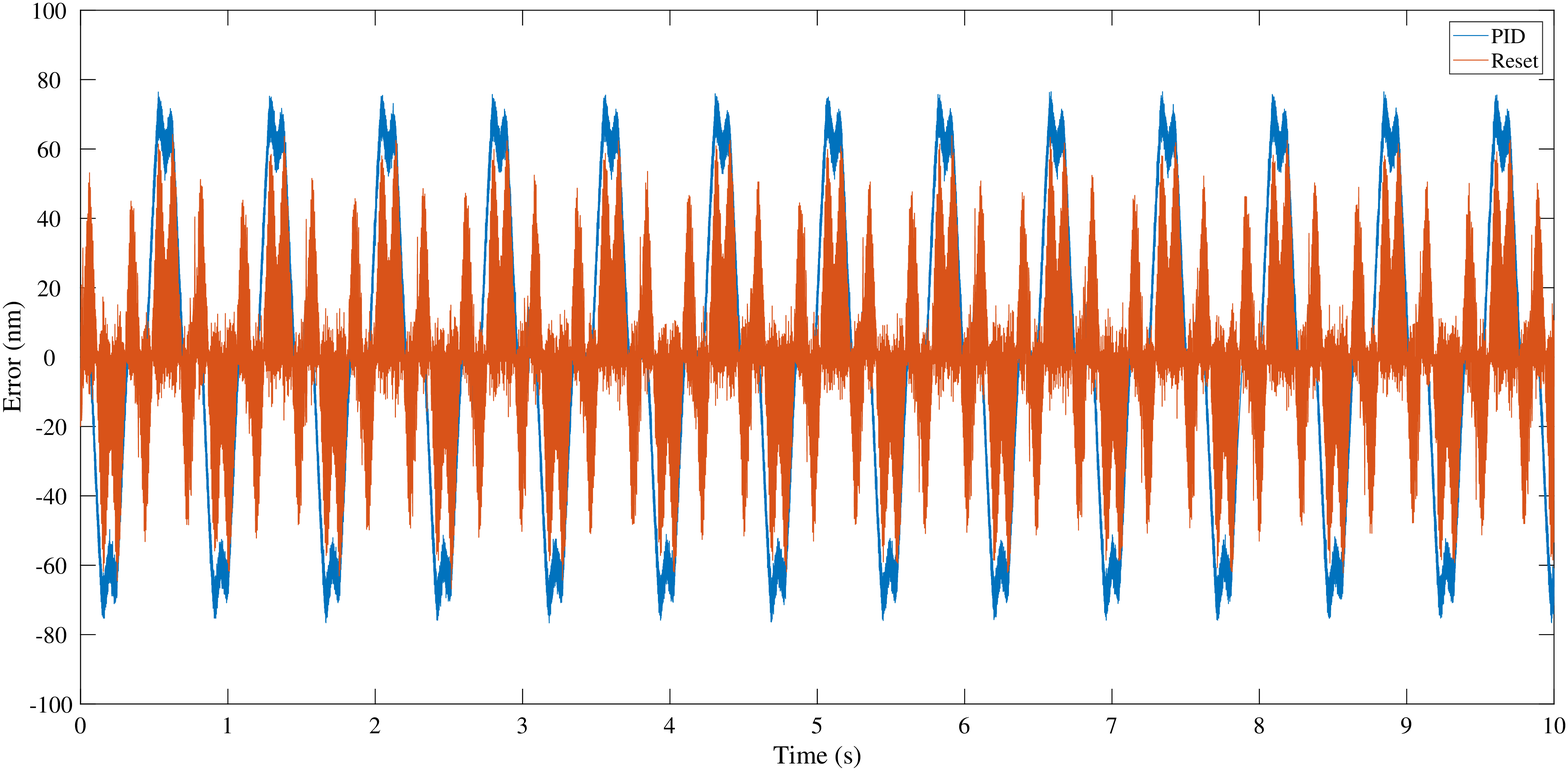}}
	\subfloat[]{\label{tr:b}\includegraphics[width=0.2405\textwidth]{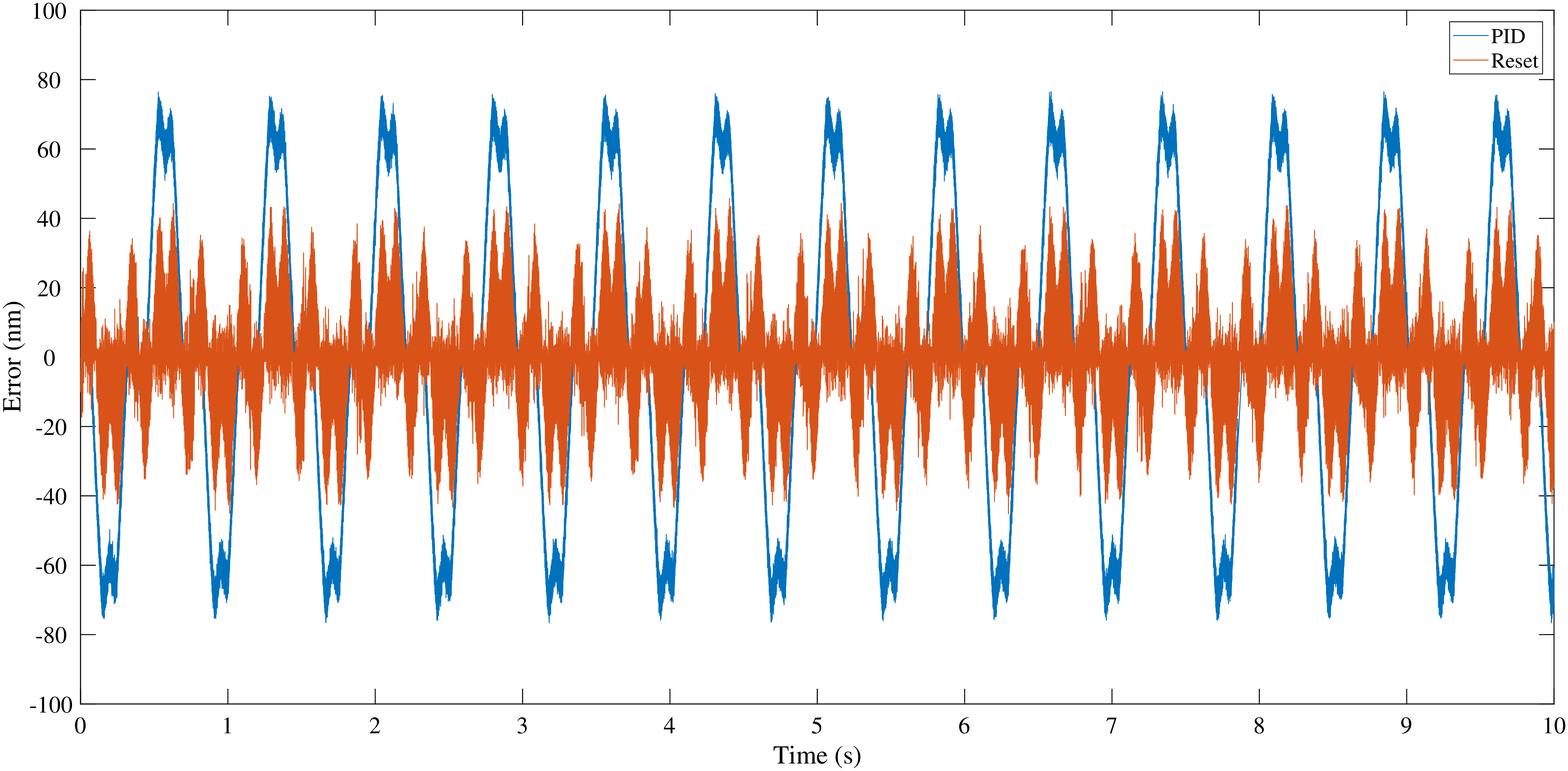}}
	\caption{\label{fig:tr} (a) Case 1 -Tracking error for Reset controller A vs PID. (b) Case 2 - Tracking error for Reset controller B vs PID. }
\end{figure}

Fourth-order trajectory planning as formulated in \cite{profile4} is used to create a triangular wave reference signal with an amplitude of $\SI{400 }{\nano \meter}$. Also, the second-order feedforward proposed in \cite{profile4} is implemented. Maximum allowed velocity, acceleration, jerk and snap of this reference signal are limited. The tracking errors obtained from the reset controllers are compared with the error from PID in Fig. \ref{fig:tr}. RMS values of tracking errors are tabulated in Table \ref{tab:err}. The RMS error for Controller A has reduced by $\SI{34}{\percent}$ compared to PID. This can be explained by the higher low frequency gain of Controller A. The reduction for Controller B is even higher, at around $\SI{60}{\percent}$. This  also is as expected, since by increasing bandwidth, the gain of Controller B at lower frequencies is greater than that of controller A.

 \begin{table}
 	\centering
 	\captionsetup{justification=centering}
 	\begin{tabular}[width=.5\textwidth]{|c|c|c|}
 		\hline
 		Controller & Max. Steady State  & RMS Tracking \\
 		 & Error(nm) & Error(nm)\\
 		\hline
 		PID & 40 & 41.48  \\
 		Controller A & 20 & 23.12 \\
 		Controller B & 40 & 15.73\\
 		\hline
 	\end{tabular}
 	\caption{Comparison of maximum steady state errors and RMS tracking errors for the developed controllers.}	
 	\label{tab:err}
 \end{table}

\subsubsection{Steady State Precision}

\begin{figure}[!h]
	\subfloat[]{\label{ss:a}\includegraphics[width=0.2395\textwidth]{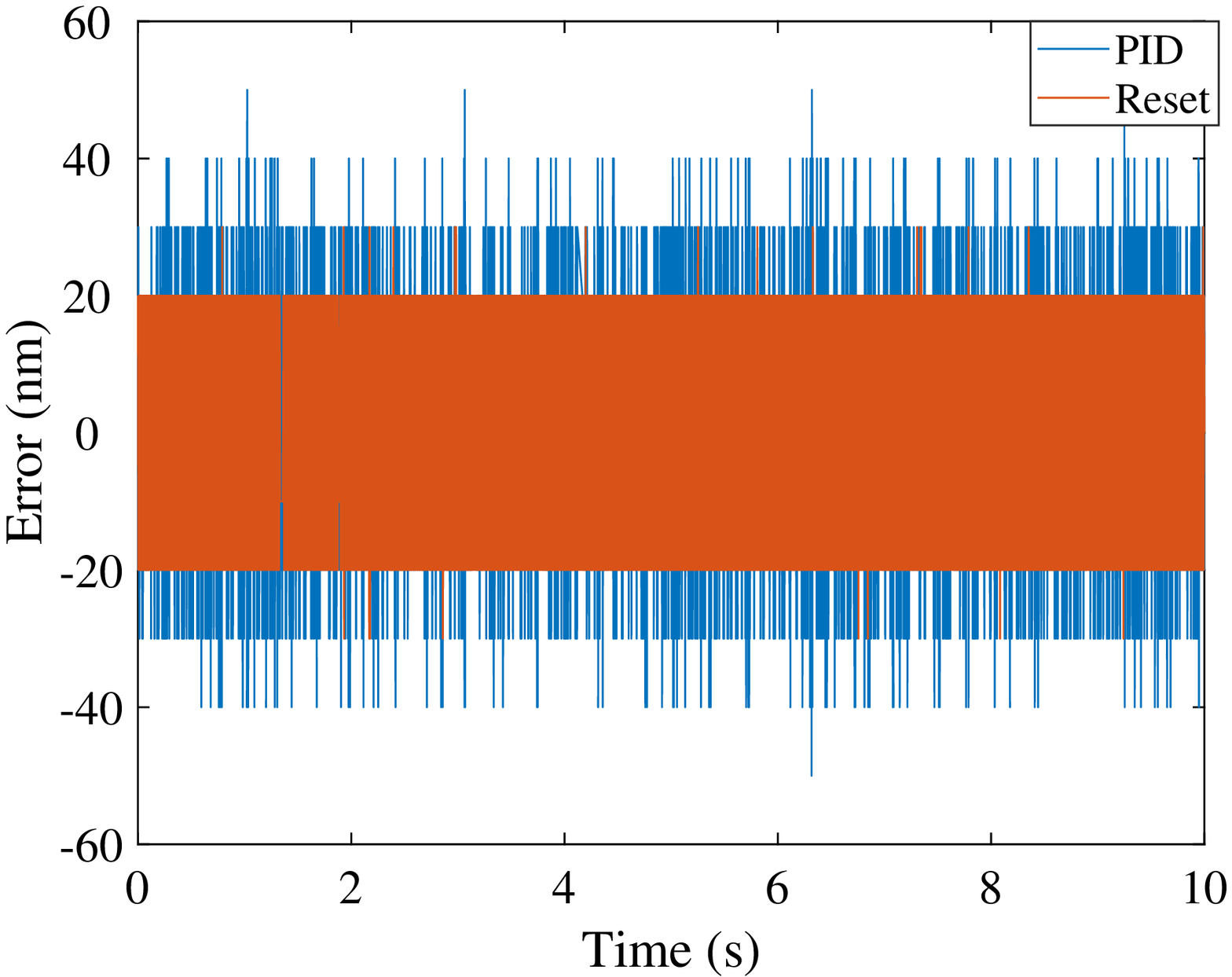}}
	\subfloat[]{\label{ss:b}\includegraphics[width=0.2405\textwidth]{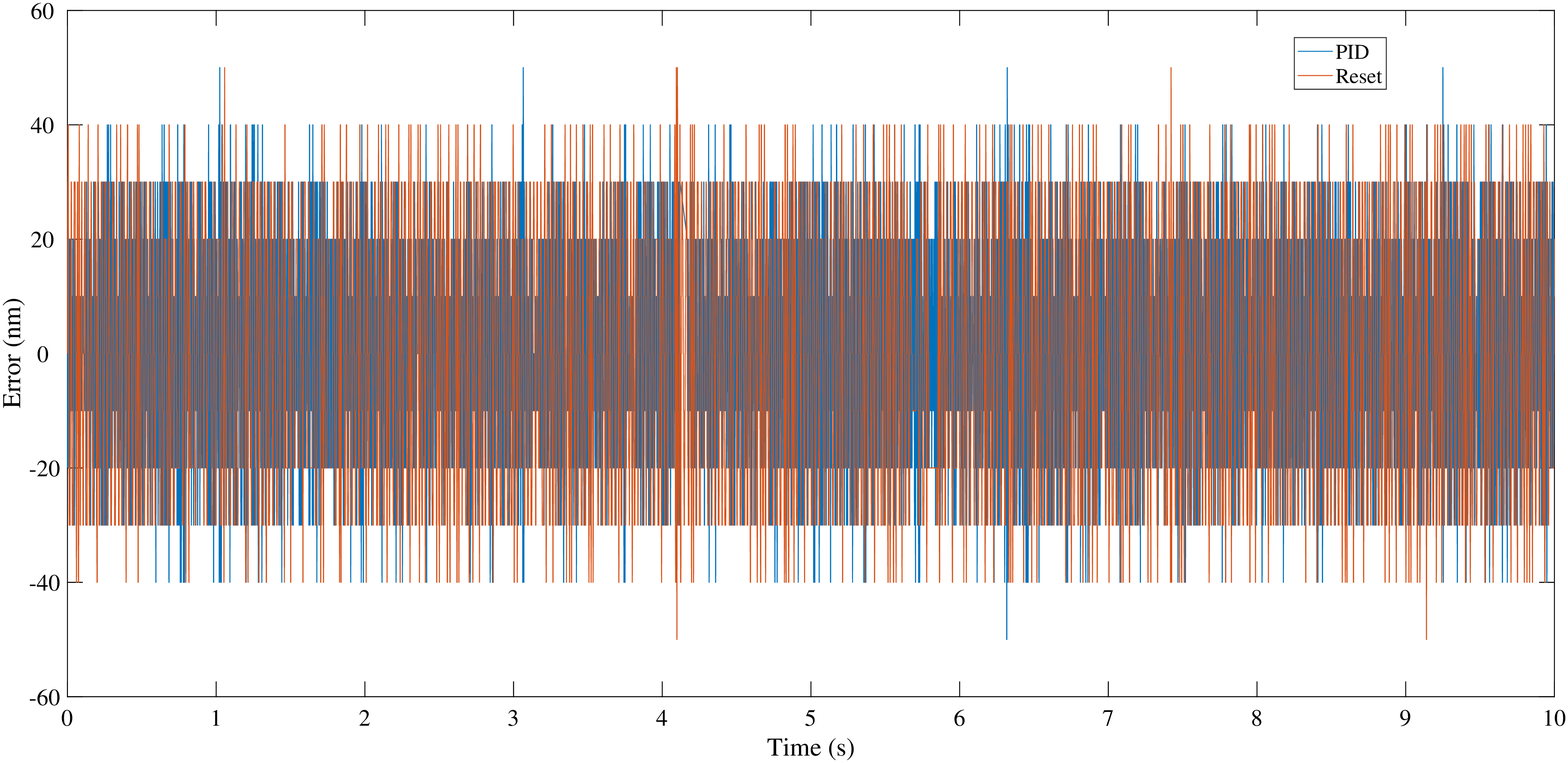}}
	\caption{\label{fig:ssp} (a) Case 1 - Steady error for Reset controller A vs PID. (b) Case 2 - Steady error for Reset controller B vs PID. }
\end{figure}

In steady state, with only sensor
noise present in the system, the achieved precision was $\SI{10}{\nano\meter}$ for all the three controllers, which is equal to the resolution of the interferometer used. Therefore, to study the improvement in precision, white noise of amplitude  $\SI{50}{\nano\meter}$ was added as measurement noise $n$  and the error signal was measured. The obtained steady state errors are compared with that of the PID in Fig. \ref{fig:ssp}, and the maximum errors are tabulated in Table \ref{tab:err}. As expected, Controller A showed reduction in steady state error while Controller B was tuned to have the same precision and hence there was no improvement.

\subsection{Frequency Domain Experiments}

\begin{figure}
	\centering
	\captionsetup{justification=centering}
	\includegraphics[width=0.3\textwidth]{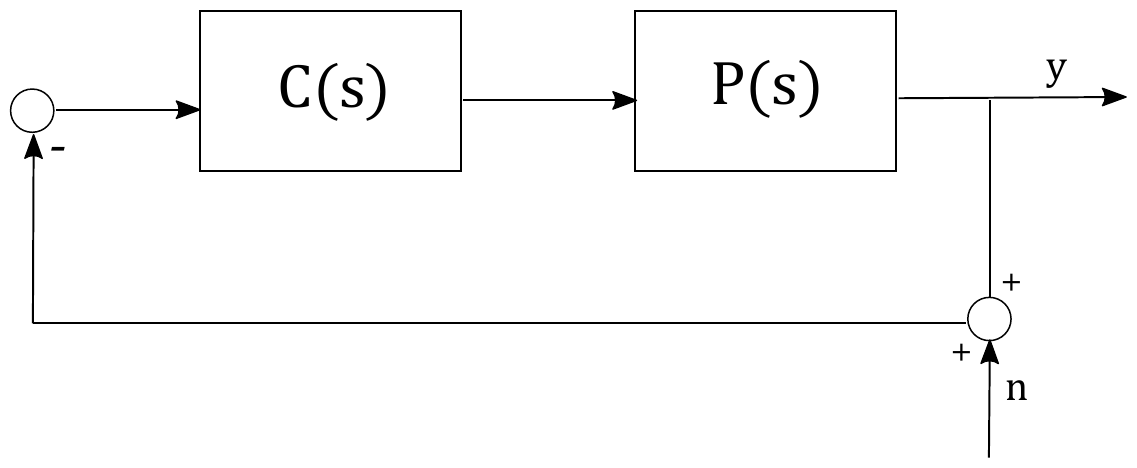}
	\caption{Block diagram of the control loop with signals $y(t)$ and $ n(t) $ used for closed loop identification. }
	\label{fig:identification_block}
	\end{figure}

In order to validate the controllers in frequency domain, a closed loop identification technique was used to compute the sensitivity function $ S(j\omega)$ and the complementary sensitivity function $ T(j\omega)$.  Both  $T(j\omega)$ and $S(j\omega)$ were identified by applying a frequency sweep at the measurement noise position $n$ in {Fig. \ref{fig:identification_block}}. $T(j\omega)$ was identified as the transfer from $-n$ to $y$, whereas $S(j\omega)$ was identified as the transfer from $n$ to $y+n$.

\begin{figure}[!h]
	\centering
	\subfloat[ ]{\label{fig:sens_rpid1_1}\includegraphics[width=0.232\textwidth]{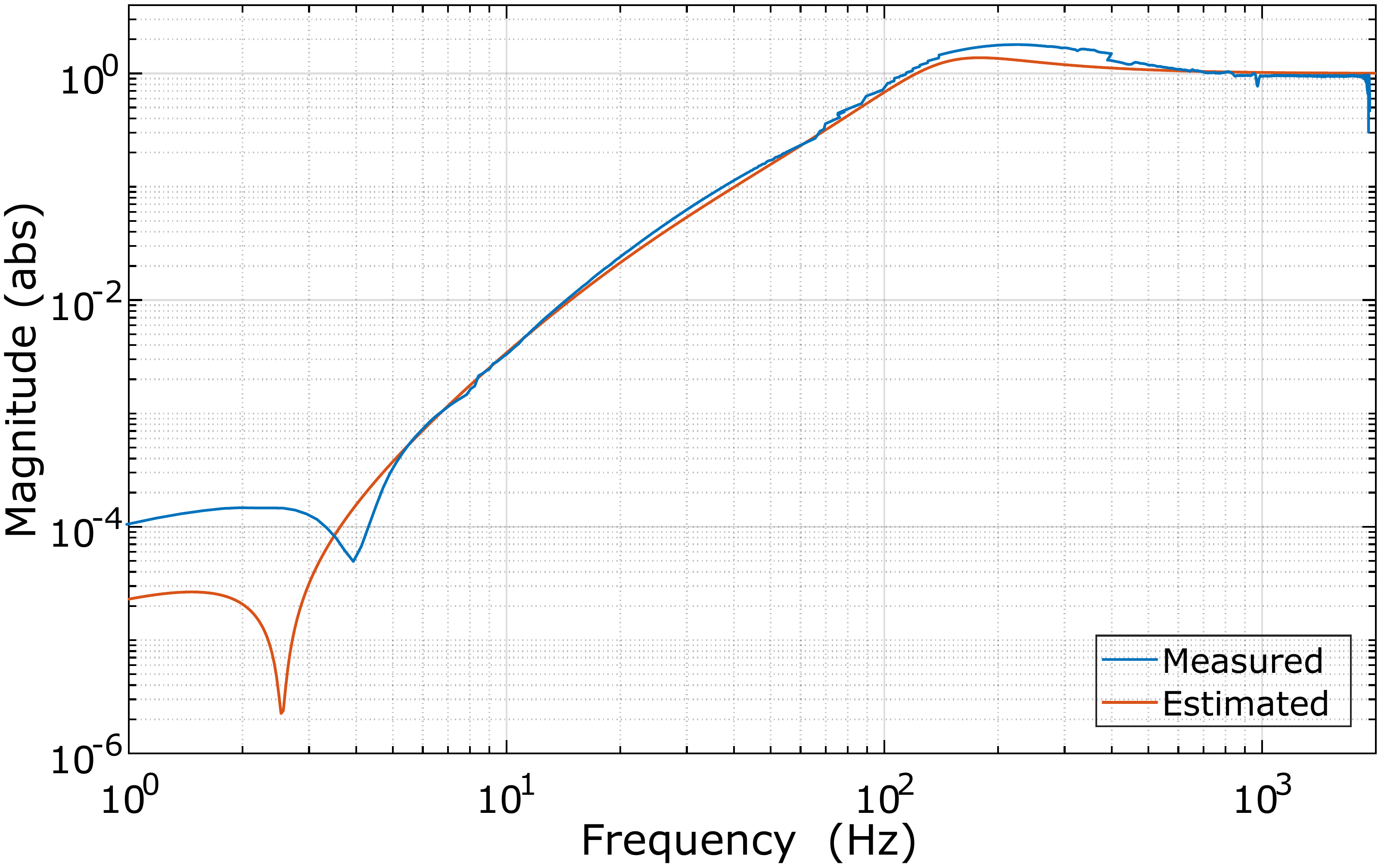}}
	\subfloat[]{\label{fig:sens_rpid1_2}\includegraphics[width=0.238\textwidth]{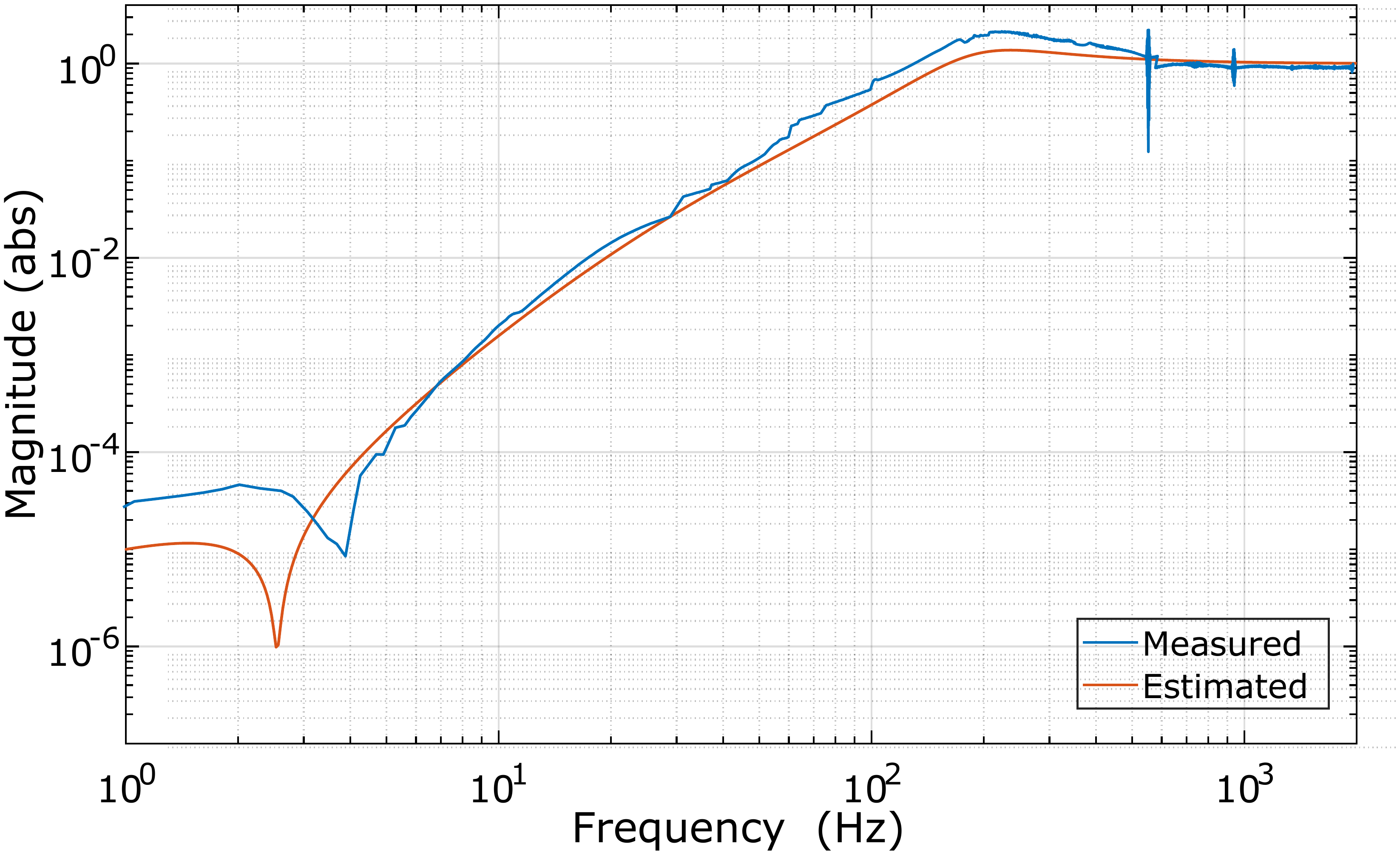}}
	\caption{\label{fig:sens} (a) Comparison of estimated and measured  sensitivity function for Controller A  (b) Comparison of estimated and measured complementary  function for Controller B  }
\end{figure}

\begin{figure}[!h]
	\centering
	\subfloat[ ]{\label{fig:compl_rpid1_1}\includegraphics[width=0.234\textwidth]{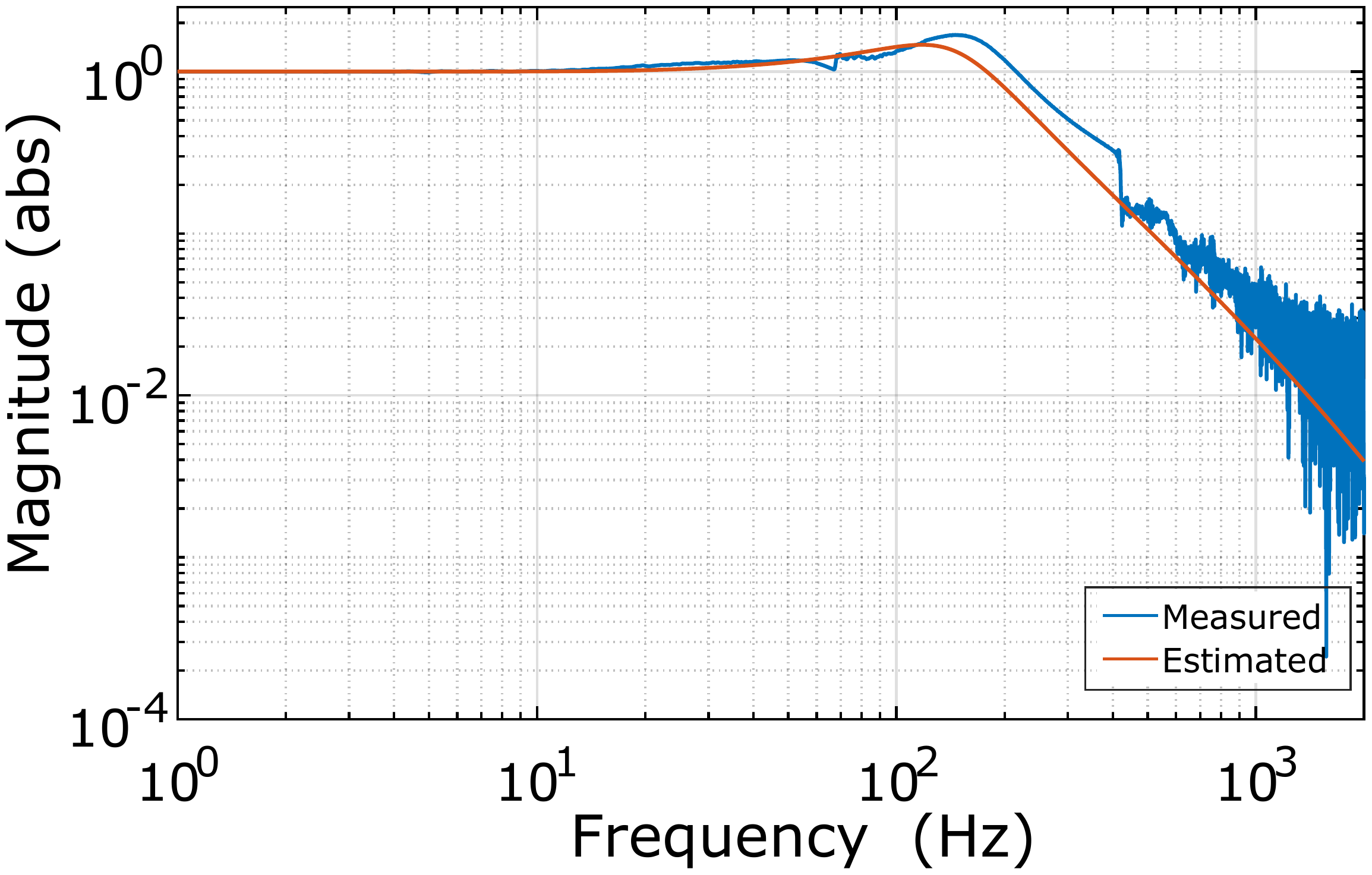}}
	\subfloat[]{\label{fig:compl_rpid2_1}\includegraphics[width=0.234\textwidth]{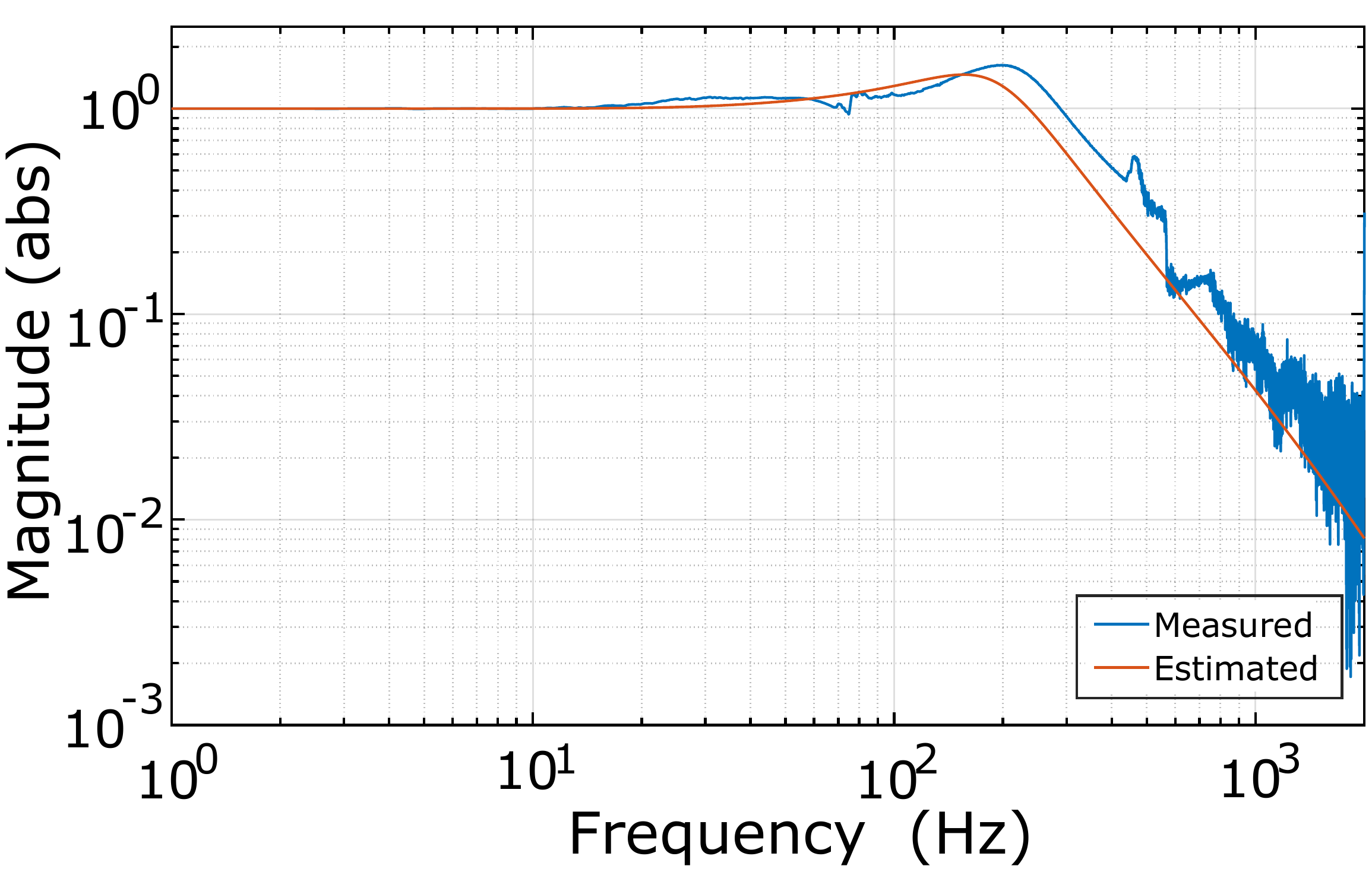}}
	\caption{\label{fig:compl} (a) Comparison of estimated and measured complementary sensitivity function for Controller A  (b) Comparison of estimated and measured complementary sensitivity function for Controller B  }
\end{figure}

The identified sensitivity functions for controllers A and B are compared with sensitivity functions estimated from describing function analysis in Fig. \ref{fig:sens}. Similarly, the complementary sensitivity functions are compared in Fig. \ref{fig:compl}. It can be seen that the measured responses match well with the estimated responses. From Fig. \ref{fig:compl_rpid2_1}, it can be seen that the bandwidth has been increased from $\SI{150}{\Hz}$ to around  $\SI{200}{\Hz}$ by Controller B, which is an improvement of  $\SI{33}{\percent}$. The sensitivity peaks in the identified responses are slightly higher than with describing functions, which shows that reset has added marginally lesser phase than anticipated from describing functions. However, describing function method has been shown to be a good approximation of phase addition by reset in reset PID controllers. 

\section{CONCLUSIONS}
 \label{sec:conc}

In this work a novel reset control method has been proposed in which reset is implemented within the framework of PID to eliminate the need for differentiating action, thus overcoming waterbed effect and achieving better performance. Two reset PID controller designs have been provided. The developed controllers were implemented in a high precision positioning setup. Controller A has been shown to provide better tracking and precision than PID, for the same bandwidth. Controller B, as expected from its higher low frequency gain, showed even better tracking and a $\SI{33}{\percent}$ improvement in bandwidth for the same precision.  Frequency response approximations made using describing function analysis are validated by identifying sensitivity and complementary sensitivity behaviour from the setup. It is shown that the identified responses match well. The fundamental trade-off between bandwidth and precision in PID has been relaxed using the novel design, and the severity of water bed effect has been reduced using reset control.



\bibliographystyle{IEEEtran}
\bibliography{reference}

\end{document}